\documentclass{sig-alternate}
\usepackage{amsmath,amsfonts,amssymb}
\usepackage{times}
\usepackage{helvet}
\usepackage{courier}
\usepackage[skip=10pt]{caption}

\usepackage{xspace,pgfplots}

\begin{document}

\title{Label Visualization and Exploration in IR}

\author {
\alignauthor Omar Alonso \\
 \affaddr{Microsoft} \\
 \email{omalonso@microsoft.com}
}

\maketitle
\begin{abstract}

There is a renaissance in visual analytics systems for data analysis and sharing, in particular, 
in the current wave of big data applications. 
We introduce RAVE, a prototype that automates the generation of an interface that uses
 facets and visualization techniques for exploring and analyzing relevance assessments data sets collected via crowdsourcing. We present a technical description of the main components and demonstrate its use. 
\end{abstract}

\noindent

\section{Introduction}

The adoption of visual analytics in the information retrieval (IR) community has been relatively low compared to other areas in computer science, like databases, that process massive large data sets. Visual analytics, a combination of techniques drawn from information visualization, data mining and statistics, allow users to directly interact with the information presented to gain insights, deduce conclusions, and make better decisions.

With the ever increasing number of data sets and emerging new data sources, the combination of automatic analysis and visual tools offers potential to gain better understanding on the many assets that the typical IR researcher, data analyst or relevance engineer has to deal with on a daily basis.  Furthermore, the inclusion of crowd data in
the form of labels that can be used for
training machine learning models or evaluating the quality of search engine results, offers new opportunities for using
visualization techniques.
 
In IR, collecting high quality document relevance assessments pairs is a crucial step for building
relevance models.  The task, which is very subjective, consists of assessing the relevance of a document to a given topic or query. The human (editor, judge, or worker) performs a visual inspection of the document and
provides a label on a particular relevance scale. Finally, label quality control mechanisms are used to produce
the final data set.

In contrast to  infographics or static visualization tools that can present different types of
metrics and summary statistics, we are interested in interactive visualization and exploration of 
preferences or labels from the
crowd in the context of a particular IR experiment that contains a human intelligence task. That is,
visual exploration of workers' preferences for certain scenarios as well as other behavioral data.
In other words, visual analytics for data exploration where humans also data providers  along with
data about their work performance.

We argue that visual analytics techniques should be part of the relevance assessment process to help gather better labels and discover potential issues in the experimental design or worker performance by looking at the entire data set~\cite{Alonso14}. Visualization tools help users in situations where seeing the structure of a data set in detail is better that seeing only a brief summary \cite{Munzner14}.

 Instead of focusing on the
visualization of search results for a query or other descriptive statistics, we are interested in  exploring the output of a relevance assessment task with the goal of 
recognizing patterns. Why do certain workers disagree on specific documents or topics? Are there any relevance cues on the presentation that may confuse a worker?  Can we identify difficult tasks?

In industrial settings, thousands of thousands of labels are collected weekly 
using data pipelines that select query document pairs and upload them to internal or external crowdsourcing platforms. The data analyst  then looks for specific metrics or anomalies in  
data sets  using
traditional database-driven tools. 
In this context,  visual analytics should allow users to analyze data when they do not know exactly what questions they need to ask in advance. 

We automate the construction of a visualization interface given the results of a crowdsourcing task.
RAVE (Relevance Assessments and Visual Exploration) is a prototype that uses as input the results of a labeling task and produces a faceted-based visualization interface for exploring and analyzing relevance assessments. We demonstrate how RAVE can be utilized to gather better insights from judges, data sets, and labels. 

\section{System Overview}

We make the following assumptions in our architecture in terms of tools and data access. Our user,
the data analyst, has access to a database of queries, topics, and documents. Human intelligence tasks
are implemented in an external crowdsourcing platform (e.g., Mechanical Turk, CrowdFlower, etc.) or 
internal equivalent tool. 

\begin{table*}[ht]
\centering
\begin{tabular}{|l|c|c|c|c|c|c|c|c|}
\hline 
Query & doc\_A & doc\_B & query length & query type &has\_entity & label & worker\_id & work time \\ \hline \hline
youtube & r1& r2& 1 &navigational &company &A& 1& 19 \\ \hline
youtube & r2& r1& 1 &navigational &company &A& 2& 7\\ \hline
youtube & r1& r2& 1 &navigational &company &A& 3& 8\\ \hline
selena gomez &r2& r1 &2& informational& person& same &1& 21\\ \hline
selena gomez &r1& r2 &2&informational &person &B& 4& 37\\ \hline
selena gomez &r2& r1 &2&informational &person &B& 3& 9\\ \hline
\end{tabular}
\caption{Data description example  for a relevance assessment task.  Columns doc\_A and doc\_B represent the content of the A-B comparison; r1 and r2 are the rankers. Query length in characters, query type, and 
has\_entity are 
query annotations. Worker\_id and work time in seconds represent assignment metadata. }
\label{table:sample}

\end{table*}

The user begins the implementation of the experiment by sampling query-document pairs $\langle q, d \rangle$ from a database.
The second step is to annotate the sampled data by running some classifiers and NLP tools such as query type identification (e.g., navigational, informational, transactional) and named-entity recognition (e.g., person, organization, 
location, etc.) to augment the original data with annotations $\langle e_{1},\dots,e_{n} \rangle$.
Once the crowdsourcing task is completed, the labels provided by workers and other assignment metadata
(e.g., worker id, time spent, approval rate, etc.), $\langle  l, a_{1},\dots,a_{m} \rangle$,
are  available. We can think of the underlying data
representation as query-document pairs with query annotations, labels, and assignment metadata. That is, 
for a single assessment, a tuple
$\langle q, d, e_{1},\dots,e_{n}, l, a_{1},\dots,a_{m} \rangle$ where {\em l} represents the label.
Table~\ref{table:sample} shows an example.

Relevance assessment is a visual exercise and capturing the image of what the worker sees at assessing time is an important part of our approach. Each document is saved as an image and the workers perform
the task looking at the same set of images. This allows our user to see the same content as workers.

\section{Relevance Assessment Visualization and Exploration}

As mentioned earlier, RAVE uses the image of the  document as the visual focus and  query annotations and assignment metadata as facets.  
The prototype automatically generates  visualizations and facets for a couple of available tools: Pivot\footnote{\url{http://bit.ly/1DsLdC6}} and Exhibit\footnote{\url{http://simile-widgets.org/exhibit/}}. 
We now describe the specific details for the automation.

\subsection{Human Intelligence Task}
As driving example, we would like to evaluate the quality of a new ranking function against an
existing baseline. That is, assess the relevance of two ranking functions $r_1$ and $r_2$, each returning a fixed number of documents 
$d_1,\ldots,d_n$ (where ${n < 10}$) as results for a query $q$. For collecting the assessments, we create 
an A-B comparison task that shows the query and the results for the two rankers in random order (a
ranker may appear in column A or B).  The task for the workers
 is to select which search results they prefer according to three choices: A is better, B is better, or they are the same as third option.

As part of the data preparation step, the tool captures a debranded SERP (Search Engine Results Page)  
screenshot for a query document pair. The document,  in this case the ranked hit list, is saved as an image using standard libraries. A debranded page means 
that there are no specific user interface  items that may bias  workers.

A bit more processing is needed for Pivot, which requires the use of the Pauthor command line tool for generating  Deep Zoom images. For Exhibit, the thumbnail version of the original image is also produced. A configuration file is needed for specifying which columns from the results data file 
corresponds to which facets.

\subsection{Generating a Pivot Collection}
Pivot collections are stored in a CXML schema that defines facets and other properties. In essence, the collection is a \texttt{cxml} file that describes the facets and contains all the elements needed for 
presentation. The generation of the collection works as follows. 
 The code first outputs the facets that we are interested in (\texttt{FacetCategory}) and then
 loops through the rows of the input file (the experiment task results)  and outputs an item for each entry (\texttt{Item}). 

The final result is an XML file with all the facets, values and images that can be visualized. To be able to see the visualization, a Silverlight plug-in or the Pivot viewer application are needed for rendering the \texttt{cxml} file. 

\begin{figure*}[h]
\centering
\includegraphics[width=17cm]{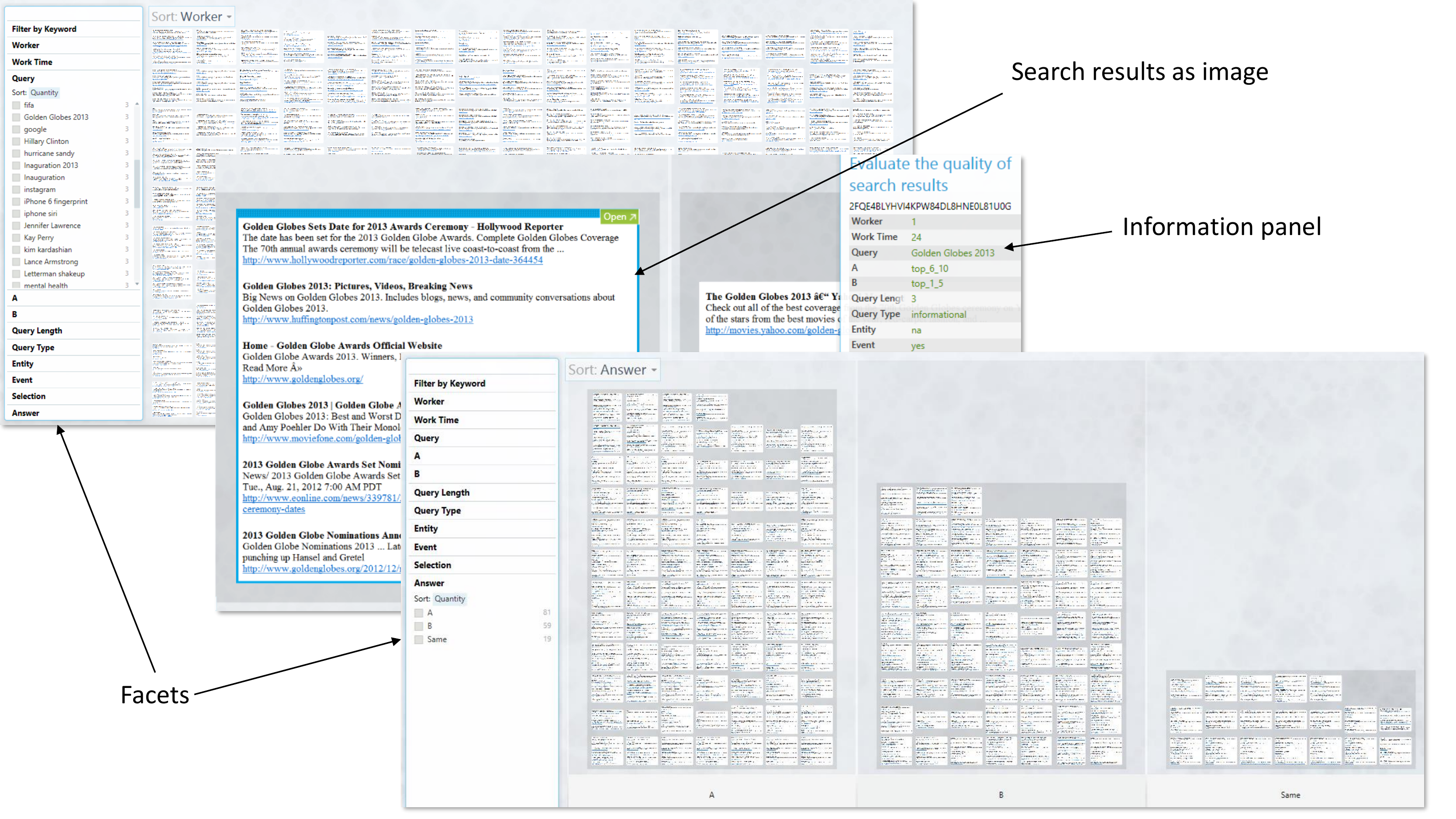}
\caption{ Pivot collection visualization. Three screenshots of the tool in action. From background to front: overview of the collection (all  images) sorted by queries, focus on the search results for the query \{Golden Globes 2013\}, distribution of ranker preferences for a data set. }
\label{fig:pivot}
\end{figure*}

Figure~\ref{fig:pivot_code} shows a snippet of the schema and collection code.  Figure~\ref{fig:pivot} shows the exploration of the results sets for a specific query. We can observe visually that ranker that 
appears in column A wins over ranker in column B and that a few judges believe that both are the same. By selecting a document
from the ``same'' answer we can investigate further why this was the case.

\begin{figure}
\small{
\begin{verbatim}
<FacetCategories>
  <FacetCategory Name="Worker" Type="String"/> 
  <FacetCategory Name="Query" Type="String"/> 
  ... 
  <FacetCategory Name="Answer" Type="String"/> 
</FacetCategories> 
<Items> 
  <Item Id="0" Name="..." Img="Selena_GomezB.png"> 
   <Description>...</Description> 
   <Facets> 
    <Facet Name="Worker">
       <String Value="4"/>
    </Facet> 
    <Facet Name="Query">
       <String Value="Selena Gomez"/>     
    </Facet> 
    <Facet Name="Answer">
      <String Value="B"/>
      </Facet> 
   </Facets> 
  </Item> 
  <Item Id="1" Name="..." Img="Selena_GomezB.png"> 
   <Description>...</Description> 
   <Facets> 
    <Facet Name="Worker"><String Value="1"/>
    </Facet> 
    <Facet Name="Query">
       <String Value="Selena Gomez"/>    
    </Facet> 
    <Facet Name="Answer"><String Value="Same"/>
     </Facet> 
   </Facets> 
</Item> ... </Items> 
\end{verbatim}
}
\caption{Snippet of the cxml code that describes the collection. }
\label{fig:pivot_code}
\end{figure}

\subsection{Generating an Exhibit}

Exhibit is implemented as an open source JavaScript library and there is no software to install; everything works on the browser. RAVE generates two files for creating an Exhibit: an HTML file that contains the layout of the elements in the web page and the data file in json format.
For producing the json view, in a similar fashion to Pivot, the prototype reads the results of the experiment  output and produces all the information needed.

\begin{figure*}[h]
\centering
\includegraphics[width=17cm]{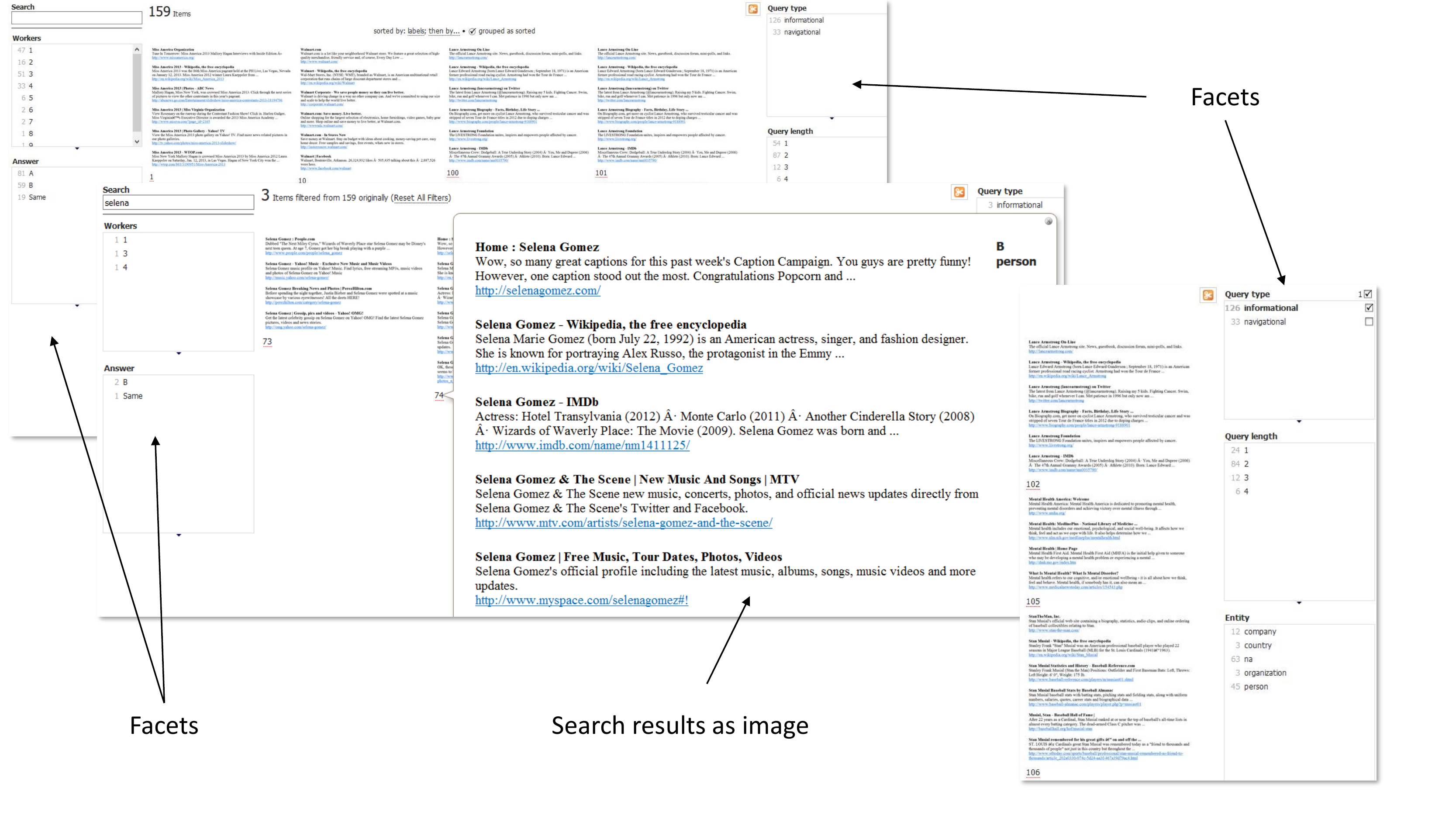}
\caption{ Exhibit collection visualization. Three screenshots of the 
tool in action. From background to front: overview of the collection with thumbnail images,  focus on search results for the query \{Selena Gomez\}, more facets on the right side of the web page.}
\label{fig:exhibit}
\end{figure*}

Figure~\ref{fig:exhibit-code} describes sample data and Figure~\ref{fig:exhibit} shows the exploration of the results sets for a specific navigational query to investigate if there was any
difference in the preferences for both rankers.

\begin{figure}
\small{
\begin{verbatim}

{ 
  types: {
  `Item': { 
  pluralLabel: `Items'
  } 
 } 
,items: [
{ type: `Item',
label: `4', worker: `4', 
querytype: `informational', 
answer: `B', 
image: `Selena_GomeztopB.png', 
thumbnail: `Selena_GomeztopB_tb.png',
},
{ type: `Item',
label: `1', worker: `1', 
querytype: `informational', 
answer: `Same',
image: `Selena_GomeztopB.png', 
thumbnail: `Selena_GomezopB_tb.png',
},
... ] }
\end{verbatim}
}
\caption{Exhibit collection source code. A json snippet that describes the data set.}
\label{fig:exhibit-code}
\end{figure}

\begin{figure}[h]
\centering
\includegraphics[width=8.8cm]{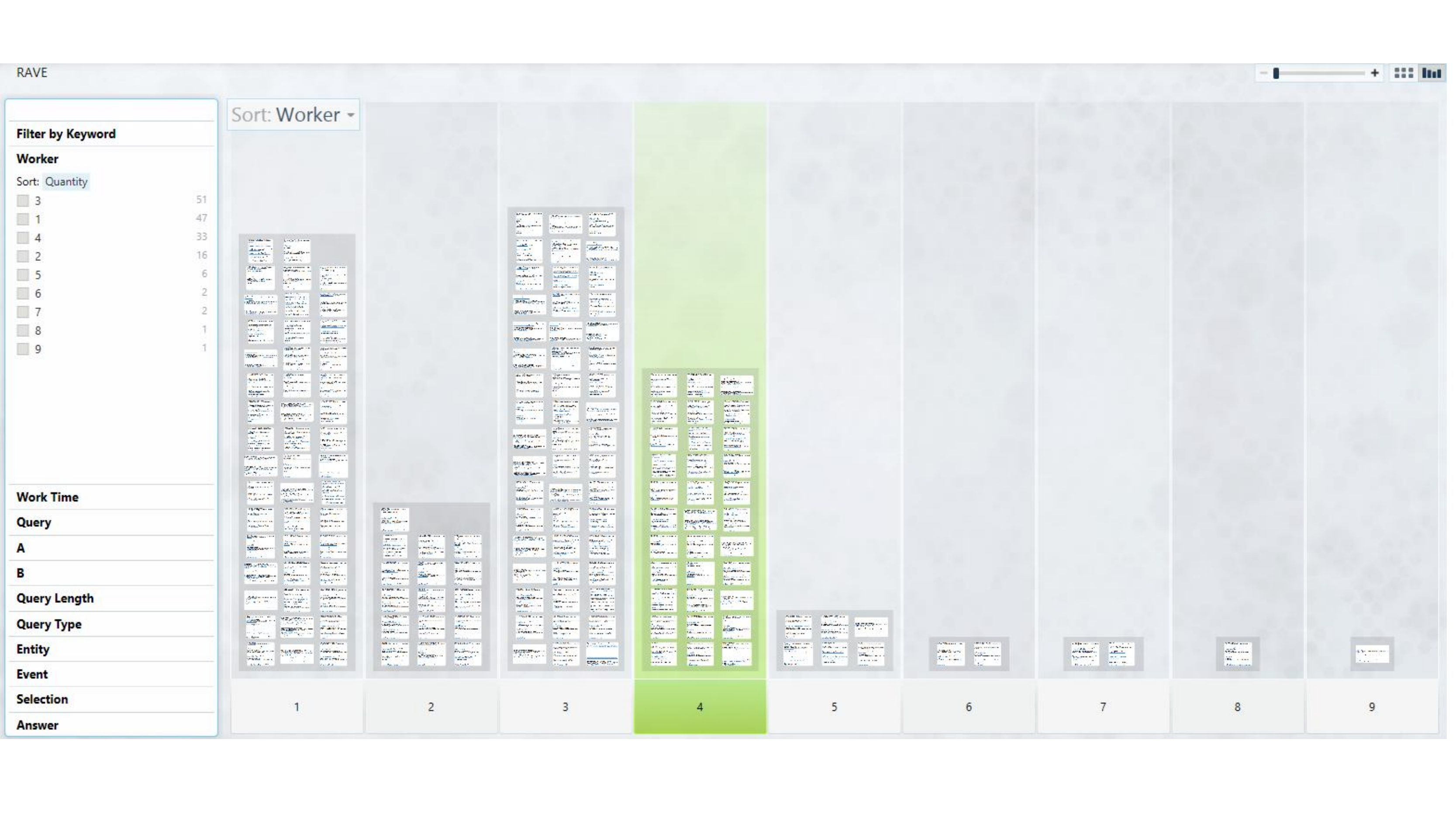}
\caption{Analyzing workers' work units by quantity.}
\label{fig:pivot-worker}
\end{figure}

\subsection{Work Data Exploration}

Now that we have shown how to visualize a data set, we can explore workers' data to identify patterns. For example, of the nine workers who participated in the task, three of them have performed most of the work and two worked on single assignments (Figure~\ref{fig:pivot-worker}). By zooming into the workload of a particular worker we can investigate his/her answers in more detail by comparing to the answer of other workers. The tool then allows the user to explore not only the final labels but also the specific worker
data that can help determine worker quality and other behaviors. 

\section{Related Work}
As researchers and practitioners collect and analyze their own labeled data sets, new tools and
solutions that facilitate such tasks are becoming available. Examples are end-to-end industrial crowdsourcing pipelines \cite{Kandylas},
the automation of crowdsourcing relevance with Terrier~\cite{McCreadieMO12}, and an open source system for collecting relevance assessments~\cite{KoopmanZ14}. 
On the visual analytics front, VIRTUE, a system for exploring IR system performance and related metrics is described in \cite{AngeliniFSS14}. 
SeeDB, a visualization recommendation engine for fast visual analysis is 
presented in~\cite{VartakMPP14}.
Finally, there is emerging work on using visualization to help collect good labels via crowdsourcing
in the NLP annotations~\cite{LiSXZ15}.

\section{Conclusion and Future Work}

We showed a prototype that can automatically generate a facet-based visualization for exploring a collection of relevance assessments collected via crowdsourcing.
While this may look like a very narrow space, in practice, practitioners spend considerable amount of time looking at labeled data before the relevance modeling phase. 
 Our goal is to assist data analysts who need to collect and assess relevance tasks labels by allowing them to visually explore those data sets
in more detail. As an example, we  showed an A-B comparison experiment but the techniques presented work for any type of task that requires workers to visually explore content and produce some label.

We are not interested in imposing a particular visualization metaphor but rather to suggest the adoption of this type of tools as part of the relevance assessment gathering process in IR.  The prototype
offers the visualization for two tools and can be extended to others. The Exhibit example is
very flexible and easy to deploy making it a low-cost development alternative. 

Visually exploring a data set can be useful to decide if the labels are of good quality, if there are
no potential issues with the experiment or if the presentation of the results can bias the final labels.
RAVE differs from previous research work in the sense that our focus is on exploring data sets instead of visualizing metrics.  
 With RAVE the user can identify patterns and perform comparisons. 

Future work includes automating the recommendation of visualizations, using the prototype
to explore  other existing assessments data sets like the 
TREC relevance labels and investigate the integration with other toolkits like D3.

\bibliographystyle{plain}
\bibliography{ref-list}
\end{document}